\documentstyle[titlepage,12pt]{article}
\begin{document}
\renewcommand{\thesection}{\Roman{section}.} \baselineskip=24pt plus 1pt
minus 1pt
\begin{titlepage}
\vspace*{0.5cm}
\begin{center}
\LARGE\bf Quantum Revivals in Periodically Driven Systems\\ close to nonlinear 
resonance
\\[1.5cm]
\normalsize\bf Farhan Saif$^{1,2,*}$ and  Mauro Fortunato$^{1,3}$
\end{center}
\vspace{7pt}
\begin{description}
\item [$^1$] Abteilung f\"ur Quantenphysik, Universit\"at Ulm,
Albert-Einstein-Allee 11, D-89069 Ulm, Germany
\item [$^2$] Department of Electronics, Quaid-i-Azam University,
54320 Islamabad, Pakistan 
\item [$^3$] Dipartimento di Matematica e Fisica and INFM
Universit\`a di Camerino, I-62032, Italy
\end{description}
\vspace{0.3cm}

\begin{center}
\normalsize 
We calculate the quantum revival time for a wave-packet initially well
localized in a one-dimensional potential in the presence of an
external periodic modulating field. The dependence of the revival time on
various parameters of the driven system is shown analytically. As an example
of application of our approach, we compare the analytically obtained values
of the revival time for various modulation strengths with the numerically
computed ones in the case of a driven gravitational cavity. We show that
they are in very good agreement.
\end{center}
\noindent Present Address: Department of Physics, the University of Arizona, 
Tucson 85721, Arizona USA.  E.mail: saif@physics.arizona.edu
\end{titlepage}
\newpage

In nature interference phenomena lead to
recurrences~\cite{kn:talbot,kn:bowo}.
In quantum mechanical evolution, for instance, interference plays
an important role and manifests itself in quantum recurrences~\cite{marz}.
A quantum wave-packet spreads all over the available space after a
few classical periods following wave mechanics and collapses. However,
due to quantum dynamics it rebuilds itself after a
certain evolution time which is larger than the classical period.
In one-dimensional systems this phenomenon is well studied and is known
as quantum revivals and fractional
revivals~\cite{kn:parker,kn:aver,kn:leic,kn:brau,kn:chen,kn:stroud}.
In higher dimensional systems, which exhibit classical and quantum
chaos,
this phenomenon has been numerically observed
earlier~\cite{kn:haak,kn:hogg,kn:bres,kn:toms,saifee}.
In particular, the quantum revivals occurring in driven gravitational
cavities~\cite{kn:saif2,kn:saif} have been investigated analytically
and
numerically.

In this Letter we provide a general analytical prescription
to the revival phenomenon occurring in any periodically-driven
time-dependent
systems. By using semi-classical secular
theory~\cite{kn:berm,kn:flat,kn:reichl}, we derive a simple relation
which allows us to calculate the quantum revival time in the presence
of the
external periodic modulation. For the sake of concreteness, we apply
our
general result to the dynamics of atoms in a modulated gravitational
cavity~\cite{kn:cohen} which is accessible to laboratory
experiments~\cite{kn:steane}. Hence we argue that dynamical revivals
are generic to all periodically-driven explicitly time-dependent
systems,
so far as resonances prevail.

Let us consider a one-dimensional system driven by an external periodic
field. In order to calculate its quasi-energy, we consider the
nonlinear
resonances of the time-dependent system~\cite{kn:berm,kn:flat}.
We denote the eigenstates and eigenvalues of the corresponding time
independent system by $|n\rangle$ and $E_n$, respectively, so that
$H_0|n\rangle=E_n|n\rangle$, where $H_0$ is the unperturbed
Hamiltonian.
We consider the evolution of a wave-packet in an arbitrary
one-dimensional
potential in the presence of an external periodic field. The
Hamiltonian of the driven system in dimensionless form can be expressed
as
\begin{equation}
H=H_0+ \lambda\,V(x)\sin t\;,
\label{eq:sche}
\end{equation}
where $\lambda$ is the dimensionless modulation strength and $V(x)$
defines the coupling.

In order to study the quantum nonlinear resonances of the system we
make the ansatz~\cite{kn:berm,kn:flat} that the solution of the
Schr\"odinger
equation corresponding to the Hamiltonian~(\ref{eq:sche}) can be
written in
the form
\begin{equation}
|\psi(t)\rangle=\sum_{n} C_n(t) |n\rangle
\exp\left\{-i\left[E_r+(n-r)\frac{k^{\hspace{-2.1mm}-}}{N}\right]
\frac{ t}{k^{\hspace{-2.1mm}-} }\right\}\;,
\label{eq:1}
\end{equation}
where $k^{\hspace{-2.1mm}-}$ denotes the effective Planck's constant,
and
$E_r$ is the average energy of the wave-packet in the $N$th resonance.

We substitute Eq.~(\ref{eq:1}) into the Schr\"odinger equation and
project the
result onto the state vector $\langle m|$. This leads to
\begin{eqnarray}
ik^{\hspace{-2.1mm}-}\dot C_m & = & \left[E_m-E_r+
(m-r)\frac{k^{\hspace{-2.1mm}-}}{N}\right] C_m(t)
+\frac{\lambda}{2i}\sum_n V_{m,n}
\nonumber\\
 & & \times \left[e^{i(n-m+N)t/N}- e^{i(n-m-N)t/N}\right]C_n(t)\;,
\label{eq:1a}
\end{eqnarray}
where $V_{n,m}=\langle n|V(x)|m \rangle$ are the matrix elements of
$V(x)$.
In Eq.~(\ref{eq:1a}) we drop the fast oscillating terms and
keep only the resonant ones, that is, $n=m\pm N$.
Moreover, for large $m$ we may take~\cite{kn:berm,kn:flat}
$V_{m,m+N}\approx V_{m,m-N}=V$. Hence, Eq.~(\ref{eq:1a}) reduces to
\begin{eqnarray}
ik^{\hspace{-2.1mm}-}\dot C_m=\left[E_m-E_r
+ (m-r)\frac{k^{\hspace{-2.1mm}-}}{N}\right] C_m(t)
\nonumber\\
+\frac{\lambda V}{2i}\left(C_{m+N}-C_{m-N}\right)\;.
\label{eq:2a}
\end{eqnarray}
In view of the slow dependence of the energy $E_m$ on the quantum
number
$m$ around the initially excited level $r$ in a nonlinear resonance, we
expand the energy $E_m$ up to second order. Hence, the equation
of motion for the probability amplitude $C_m$ takes the form
\begin{eqnarray}
ik^{\hspace{-2.1mm}-}\dot C_m&=& (m-r)\left( E'_r -
\frac{k^{\hspace{-2.1mm}-}}{N} \right)C_m
\nonumber\\
&+& \frac{1}{2}(m-r)^2 E''_r C_m
+\frac{\lambda V}{2i}\left(C_{m+N}-C_{m-N}\right)\;.
\label{eq:3}
\end{eqnarray}
For the exact resonance case $E'_r=k^{\hspace{-2.1mm}-}/N$, and as a
result the first term on the right hand side vanishes. We may write
this equation in angle representation by introducing the Fourier
representation of $C_m$ as
\begin{eqnarray}
C_m&=&\frac{1}{2\pi}\int\limits_{0}^{2\pi} g(\varphi)
e^{-i(m-r)\varphi}\,d\varphi \nonumber\\
&=&\frac{1}{2N\pi}\int\limits_{0}^{2N\pi} g(\theta)
e^{-i(m-r)\theta/N}\,d\theta\;.
\label{eq:3a}
\end{eqnarray}
This particular choice of the Fourier representation of $C_m$
yields a Schr\"odinger-like equation,
$ik^{\hspace{-2.1mm}-}{\dot g}(\theta,t)={\cal H}g(\theta,t)$,
where the Hamiltonian ${\cal H}(\theta)$ reads
\begin{equation}
{\cal H}(\theta)=-\frac{N^2E''_r}{2}\frac{\partial^2}{\partial\theta^2}
+\lambda V\sin(\theta)\;,
\label{eq:4}
\end{equation}
and is independent of time. To obtain this equation, we have assumed
that
the function $g(\theta,t)$ has a $2N\pi$-periodicity in
$\theta$-coordinate.

Due to the time-independent behavior of ${\cal H}(\theta)$, we can
write the time evolution of $g$ as $g(\theta,t)={\tilde g}(\theta)
e^{-i{\cal E}t/k^{\hspace{-1.7mm}-}}$. On substituting this
representation
in the equation of motion for $g(\theta,t)$, we are left with the
rather
simple equation
\begin{equation}
\left[-\frac{N^2 E''_r}{2}\frac{\partial^2}{\partial\theta^2}
-{\cal E} +\lambda V\sin(\theta)\right]
\tilde{g}(\theta)=0\;.
\label{eq:5}
\end{equation}
With the change of variable $\theta=2z+\pi/2$, we can reduce it to the
standard
Mathieu equation
\begin{equation}
\left[\frac{\partial^2}{\partial z^2}
+a-2q\cos(2 z)\right]
\tilde{g}(z)=0\;.
\label{eq:6}
\end{equation}
Here, $a=8{\cal E}/N^2 E''_r$, $q=4\lambda V/N^2 E''_r$, and
$\tilde{g}(\varphi)$ is a $\pi$-periodic function.

Through this procedure we have mapped our initially unsolvable
time-dependent Schr\"odinger equation onto the known Mathieu
equation~\cite{kn:berm,kn:flat} under the assumption of exact
resonance.
The Floquet solution of the Mathieu equation~\cite{kn:reichl} can be
written
as $\tilde{g}_\nu(z)=e^{iz\nu}P_{\nu}(z)$, where $P_\nu$ are the
Matthieu
functions~\cite{kn:abra}. Since we require a $\pi$-periodicity of
${\tilde g}$, we are forced to take only the even values of the index,
{\it i.e.}, $\nu=2k/N$.

Hence, the Floquet quasi-energy eigenvectors~\cite{kn:reichl} can be
written as
\begin{equation}
|\psi_k(t)\rangle=e^{-i{\cal E}_k
t/k^{\hspace{-1.7mm}-}}\,|u_k(t)\rangle\;,
\label{eq:qv}
\end{equation}
where ${\cal E}_k$ and $|u_k(t)\rangle$ are defined as
\begin{eqnarray}
{\cal E}_k&\equiv&\frac{N^2E''_r}{8}a_{\nu}(q)\;,
\label{eq:eneg} \\
|u_k(t)\rangle&\equiv&\frac{1}{2\pi}\sum_n
e^{int/N}\int\limits_{0}^{2\pi}
d\varphi e^{i\nu N\varphi/2} e^{-i(n-r)\varphi/2}
\nonumber \\
 & & \times P_{\nu(k)}\,|n\rangle \;.
\label{eq:amp}
\end{eqnarray}
In this way, we have obtained an approximate solution for a nonlinear
resonance of our explicitly time-dependent system.

In order to check the correctness of our result, we study the case of
zero
modulation strength, that is $\lambda=0$, for which $q=0$. The
corresponding
value for the Mathieu characteristic parameter becomes
$a_{\nu(n)}(q=0)=4(n-r)^2/N^2$. This reduces the quasi-energy given in
Eq.~(\ref{eq:eneg}) to
\begin{equation}
{\cal E}_n(q=0)=\frac{E''_r}{2}(n-r)^2\;,
\end{equation}
which is indeed the energy of the un-modulated system expanded around
$n=r$
for the exact resonance case. This is obtained by considering $k=n-r$,
and
therefore we may express $\nu_n(q=0)=2(n-r)/N$.

The eigenfunctions given in Eq.~(\ref{eq:qv}) form a complete set of
basis
vectors. Therefore, we can write the propagated wave-packet at any
later time
in terms of the quasi-energy eigenstates, such that
\begin{equation}
|\psi(z,t)\rangle =
\sum_n \xi_n e^{-i{\cal E}_nt/k^{\hspace{-1.7mm}-}}|u_n(t)\rangle\;,
\end{equation}
where $\xi_n$ describes the probability amplitude in the $n$th state.

In order to investigate the dynamical revivals, we calculate the
auto-correlation function between the initially excited wave-packet and
the
wave-packet $|\psi(z,t)\rangle$ after a certain evolution time $t$,
viz.
\begin{equation}
\langle \psi(0)|\psi(t)\rangle=\sum_n |\xi_n|^2
e^{-i {\cal E}_nt/k^{\hspace{-1.7mm}-}}\;.
\label{eq:auto}
\end{equation}
We take into account that the wave-packet is initially centered
around $n=r$ in the energy representation. Therefore, by expanding the
quasi-energy of the system around the resonant level $r$, we can write
\begin{eqnarray}
\langle \psi(0)|\psi(t) \rangle = \sum_n |\xi_n|^2
\exp\left\{-i\left[\omega^{(0)} +(n-r)\omega^{(1)}\right.\right.
\nonumber\\
\left.\left.+(n-r)^2\omega^{(2)}+\cdots\right]t \right\}\;,
\label{eq:60}
\end{eqnarray}
where $\omega^{(j)}\equiv (j!\, k^{\hspace{-2.1mm}-} )^{-1}
{\cal E}^{(j)}_{n}|_{n=r}$ and ${\cal E}^{(j)}_{n}|_{n=r}$ denotes the
$j$th derivative of ${\cal E}_n$ calculated at $n=r$.

We then compare the auto-correlation function for the driven system,
as given in Eq.~(\ref{eq:60}), with the auto-correlation function for
the
corresponding one-dimensional unperturbed
system~\cite{kn:aver,kn:saif2}. This
comparison helps us to identify the classical period for the driven
system as
\begin{equation}
T_{cl}\equiv\frac{2\pi}{\omega^{(1)}} =
\frac{2\pi k^{\hspace{-2.1mm}-}}{{\cal E}^{(1)}_n|_{n=r}}\;,
\end{equation}
and the quantum revival time as
\begin{equation}
T_{\lambda}\equiv\frac{2\pi}{\omega^{(2)}}
=\frac{4\pi k^{\hspace{-2.1mm}-}}{{\cal E}^{(2)}_n|_{n=r}}\;.
\label{eq:tlam}
\end{equation}

Let us now consider the case of small perturbations, {\it i.e.}
$\lambda \ll 1$. We may therefore use the expansion for the Mathieu
characteristic parameter $a_\nu$~\cite{kn:abra} only up to second order
in $q$. Then Eq.~(\ref{eq:eneg}) becomes
\begin{equation}
{\cal E}_n = \frac{E''_r}{8}\left(\nu_n^2
+ \frac{q^2}{2}\frac{1}{\nu_n^2- 1}\right)
+ O(q^4)\;,
\label{eq:quasen}
\end{equation}
and a simple expression for the revival time in the driven potential
can be derived from Eqs.~(\ref{eq:tlam}) and (\ref{eq:quasen}).

In general, the initially excited wave-packet is away from the center
of the
resonance and the condition $E'_r = k^{\hspace{-2.1mm}-}/N$ is not
satisfied.
For the sake of clarity, in the above discussion we have only
considered the
case of the exact resonance. However, taking into account the
non-resonant
situation, the revival time for a time-dependent system can be
calculated in
the general case and reads~\cite{kn:saif2}
\begin{equation}
T_{\lambda}=T_0\left[ 1-\frac{1}{2}\left(\frac{\lambda V}
{E_r''} \right)^2  \left(\frac{2}{N}\right)^4
\frac{3\nu_r^2+1}{(\nu_r^2-1)^3}\right]\;.
\label{eq:lm}
\end{equation}
Here, $T_0=4\pi k^{\hspace{-2.1mm}-}/E''_r$ denotes the revival time of
the
initial wave-packet in the un-driven potential and the index $\nu_n$ is
defined
as
\begin{equation}
\nu_n=\frac{2}{N}(n-r)+\frac{2(E_r'-k^{\hspace{-2.1mm}-}/N)}{NE_r''}\;,
\label{eq:unknu}
\end{equation}
where the second term on the right hand side is arising due to the
non-resonant
case. Substituting the value of $\nu_n$ from Eq.~(\ref{eq:unknu})
calculated
at $n=r$ in Eq.~(\ref{eq:lm}), we obtain the general expression for the
quantum revival time in a periodically driven system as
\begin{equation}
T_{\lambda}=T_0\left[ 1-\frac{1}{2}\left(\frac{\lambda V}
{E_r''} \right)^2 \frac{3\mu^2+N^2/4}{(\mu^2-N^2/4)^3}\right]\;,
\label{eq:lm1}
\end{equation}
where $\mu= \frac{(E_r'-k^{\hspace{-1.7mm}-}/N)}{E_r''}$.
Equation~(\ref{eq:lm1}) constitutes our main result: it expresses the
modification
of the quantum revival time in the presence of an external periodic
driving
field as a function of the modulation strength $\lambda$ and of the
other
characteristic parameters of the system.

The present approach is valid for small modulation strengths. In the weak
binding potentials, for which the level spacing decreases as we increase
the quantum number, the nonlinear resonances disappear after small modulations,
as observed numerically for gravitational cavity~\cite{kn:wallis,kn:dowling}
and also for hydrogen atoms. 
As a consequence, this technique is
reasonably good for these kind of potentials. For tight binding potentials,
for which the level spacing increases as we go up in the energy, we find nonlinear
resonances for higher modulation as well. In this kind of potential 
we need to consider higher-order terms in the expansion of the 
Mathieu characteristic parameter in order to
correctly calculate the quantum recurrence time.

As an application of our method, we consider the dynamics of
atoms in a gravitational cavity~\cite{kn:cohen} in the presence of an
external periodic field~\cite{kn:saif3}. In this case the atoms move in
a potential $x+V_0 e^{-\kappa x}$, where the linear term is due to the
gravitational potential and the exponential term comes from the
evanescent-wave field which can be obtained by total internal
reflection of
a laser light field on the surface of a glass
prism~\cite{kn:cohen,kn:steane,kn:cook,kn:baly,kn:chu,kn:grimm}.
In the approximation of a triangular well potential, the coupling
constant
for the $N$th resonance is given by
\begin{equation}
V\equiv\langle m|x|m\pm N \rangle=
- \frac{2E_m}{N^2\pi^2 [1\mp N/6m+O(m^{-2})]^2}\;,
\label{eq:definev}
\end{equation}
which in the limit of large $m$ becomes $V\cong -2E_N/N^2\pi^2$.
Substituting this expression into Eq.~(\ref{eq:lm1}), it is possible to
calculate the time of revival for a wave packet initially excited
with average energy $E_r$ around the $N$th resonance, which reads
\begin{equation}
T_{\lambda}=
T_0\left\{ 1-\frac{1}{8}\left(
    \frac{\lambda}{E_r} \right)^{2}
    \frac{3(1-\alpha)^2+a^2}{[(1-\alpha)^2-a^2]^3}\right\}\;,
\label{eq:rtspt}
\end{equation}
where $\alpha\equiv(E_N/E_r)^{1/2}$ and
$a\equiv \alpha^2 k^{\hspace{-2.1mm}-}/4E_r$.
If the initial energy is large enough, that is $E_r\gg 1$, we may
neglect $a^2$ with respect to $(1-\alpha)^2$ in Eq.~(\ref{eq:rtspt}),
obtaining the simpler formula
\begin{equation}
T_{\lambda}=T_0\left[ 1-\frac{3}{8}\left(
\frac{\lambda}{E_r} \right)^{2}
\frac{1}{(1-\alpha)^4}\right]\;.
\label{eq:rtsptn}
\end{equation}
This expression allows us to compare the analytically calculated
revival time
with the numerically computed ones.

We may prepare a wave packet by trapping and cooling cesium atoms in a
MOT~\cite{kn:cohen,kn:steane,kn:baly}.
In laboratory experiments the driving frequency can be changed from 0 to 2MHz,
see Ref.~\cite{kn:steane}. Hence if we select a frequency of
$\omega= 2\pi$x$0.93KHz$ out of this domain, for cesium atoms of mass
$M=2.2$x$10^{-25}Kg$, the dimensionless Plank's constant is $k^{\hspace{-1.7mm}-}=1$. Moreover,
by using effective Rabi frequency $\Omega_{eff}=2\pi$x$3.72KHz$ and the steepness
$0.57 \mu m$, we have $V=1$ and $\kappa=1$. In our numerical calculations we
use these values of the dimensionless parameters. In order to measure
the quantum recurrence time, we place the atomic wave packet at
$z_0=29.8\mu m$ for which $E_r=104.1$, and
$z_0=20.1\mu m$, for which $E_r=70.28$
above the atomic mirror.

Our numerical results show a complete qualitative and quantitative
agreement with the analytical results. In fact, from
Eq.~(\ref{eq:rtsptn})
we learn that ({\it i}) the revival time changes quadratically as a
function
of the strength $\lambda$ of the external modulation field, and
({\it ii}) the revival time depends inversely on the
square of the initial average energy of the wave-packet.
Our numerical investigation confirms both these analytical results.

\begin{figure}
\vspace{0.2truecm}
\caption{A comparison between the numerically computed ratio
$T_{\lambda}/T_0$ for two different initial conditions, $E_r=104.1$
(solid line with circles) and $E_r=70.28$ (solid line with squares),
and the
corresponding [Eq.~(\protect\ref{eq:rtsptn})] analytical results
(dashed lines)
in the case of an atom bouncing in a modulated gravitational cavity.
The other
parameters are $k^{\hspace{-1.9mm}-}=1$, $V_0=1$, and $\kappa=1$.
}
\label{fg:revival}
\end{figure}
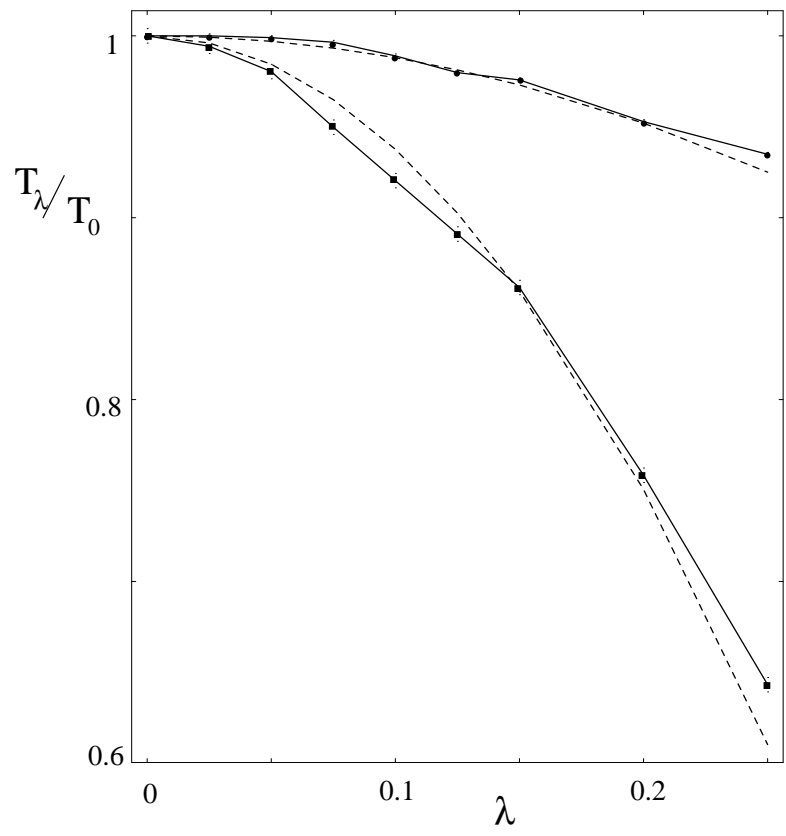

In Fig.~\ref{fg:revival} we compare the quantum revival times
calculated from
Eq.~(\ref{eq:rtsptn}) (dashed lines) with those obtained numerically
(solid
lines) for an atomic wave-packet bouncing in a modulated gravitational
cavity.
We see from the numerical data that the revival time displays a
quadratic
dependence on the strength $\lambda$ of the external modulation.
Moreover,
the change in the revival time is just about 8\% for $\lambda$ ranging
from
0 to 0.25 when the initial average energy $E_r=104.1$, (solid line with
circles), whereas it increases to almost $40\%$ for the same range of
$\lambda$ when the initial average energy $E_r=70.28$ (solid line with
squares), in agreement with the inverse square dependence of
$T_\lambda$
on $E_r$.

In summary, we have presented a general approach to the investigation
of the
phenomenon of quantum revivals in periodically driven systems. We have
derived
a simple relation which provides the quantum revival time as a
function of the modulation strength. We have finally applied our theory
to the
case of the revival time arising in the dynamics of atoms bouncing in a
modulated gravitational cavity, obtaining an excellent agreement with
the
exact numerical data. For the set of parameters which we have used in
our
calculations a modulated gravitational cavity can be realized within
the
framework of presently available technology~\cite{kn:saif2}.
Therefore we suggest an experiment, designed with the help of these
parameters,
to test the quantum recurrences in the driven gravitational cavity.

We thank G.~Alber, I.~Bialynicki-Birula, R.~Grimm, W.~P.~Schleich,
and F.~Steiner for many fruitful discussions. We gratefully thank Prof. Wolfgang
Schleich
for his hospitality at Ulm, where part of the work was carried out. 
FS acknowledges the support from QAU, and
MF
acknowledges financial contributions from INFM, EU (through the human potential 
programme "QUEST"), MURST and the university of Camerino ("Progetto Giovani Ricercatori").


\vspace{2ex}
\end{document}